\documentclass{article}
\usepackage{amssymb}
\usepackage{graphicx}
\usepackage{amsmath}

\input{tcilatex}

\begin{document}

\begin{center}
{\LARGE Universal Signature of Non-Quantum Systems}

\bigskip\bigskip

\bigskip Antony Valentini\footnote{%
email: avalentini@perimeterinstitute.ca}

\bigskip

\bigskip

\textit{Perimeter Institute for Theoretical Physics, 35 King Street North,
Waterloo, Ontario N2J 2W9, Canada.\footnote{%
Corresponding address.}}

\textit{Theoretical Physics Group, Blackett Laboratory, Imperial College,
Prince Consort Road, London SW7 2BZ, England.}

\textit{Augustus College, 14 Augustus Road, London SW19 6LN, England.%
\footnote{%
Permanent address.}}

\bigskip
\end{center}

\bigskip

\bigskip

\bigskip

It is shown that `non-quantum systems', with anomalous statistical
properties, would carry a distinctive experimental signature. Such systems
can exist in deterministic hidden-variables theories (such as the pilot-wave
theory of de Broglie and Bohm). The signature consists of non-additive
expectations for non-commuting observables, breaking the sinusoidal
modulation of quantum probabilities for two-state systems. This effect is
independent of the quantum state (pure or mixed), or of the details of the
hidden-variables model. Experiments are proposed, testing polarisation
probabilities for single photons.

\bigskip

\bigskip

\bigskip

\bigskip

\bigskip

\bigskip

\bigskip

\bigskip

\bigskip

\bigskip

\bigskip

\bigskip

\bigskip

\bigskip

\bigskip

\bigskip

\bigskip

\bigskip

\bigskip

\bigskip

\bigskip

\bigskip

\bigskip

\bigskip

\bigskip

\section{Introduction}

In a deterministic hidden-variables theory -- such as the pilot-wave theory
of de Broglie and Bohm [1--10] -- the outcomes of single quantum experiments
are determined by initial parameters $\lambda $ which, at present, we are
unable to fully control. Over an ensemble of similar experiments, these
parameters (or hidden variables) will have a distribution $\rho (\lambda )$,
giving rise to a statistical distribution of quantum outcomes. For some
particular $\rho (\lambda )$, the distribution of outcomes will agree with
the predictions of quantum theory.

It has been argued that, in pilot-wave theory and in any deterministic
hidden-variables theory, quantum physics must be viewed not only as
phenomenological but also as contingent [5, 10--18]. Specifically, quantum
theory is an effective description of a particular `quantum equilibrium'
distribution $\rho (\lambda )=\rho _{eq}(\lambda )$ of hidden variables --
much as the theory of thermal fluctuations in a classical gas is an
effective description of a thermal equilibrium distribution of molecules.
And, just as there is a wider physics of gases in thermal nonequilibrium, so
there should be -- at least in principle -- a wider physics beyond quantum
theory, of systems in `quantum nonequilibrium' $\rho (\lambda )\neq \rho
_{eq}(\lambda )$. Such systems might exist as relic particles from the early
universe \cite{10,13,14,17,18}. Or, the equilibrium state might be unstable
at high energies or in strong gravitational fields \cite{18}.

The potential uses of nonequilibrium or non-quantum systems have been
explored \cite{19}. But how might such systems be recognised experimentally?

In general, nonequilibrium systems would generate non-quantum probabilities
-- for example, anomalous blurring in single-particle interference. Quantum
interference experiments have been suggested with particles from the early
universe or from very distant astrophysical sources \cite{13,14}. However,
even if anomalous blurring were to be observed, the effect might be
explained in another way, for example by the finite size of the source%
\footnote{%
Fringe blurring is used in stellar interferometry to infer the (angular)
size of the source.} or by some peculiarity of the incoming state.

What is needed, then, is a distinctive signature of quantum nonequilibrium,
that could not be explained by some other effect involving ordinary physics.
The aim of this paper is to describe such a signature.

Our signature consists of non-additive expectations for non-commuting
observables. This would manifest as a breakdown of the sinusoidal modulation
of quantum probabilities for two-state systems, leading for example to
anomalous polarisation probabilities for single photons. As we shall see,
this signature has the attractive features of being independent of the
quantum state, and independent of the details of the hidden-variables theory.

We emphasise that our argument is quite general. It applies to any system
with a two-dimensional subspace.

\section{Additive Expectations}

In quantum theory, it is impossible to devise an experiment that measures
non-commuting observables $\hat{\Omega}_{1}$, $\hat{\Omega}_{2}$
simultaneously. The measurements are `incompatible'. However, it is possible
to devise an experiment $E_{1}$ that measures $\hat{\Omega}_{1}$ and another
experiment $E_{2}$ that measures $\hat{\Omega}_{2}$, and a third experiment $%
E$ that measures a linear combination $\alpha _{1}\hat{\Omega}_{1}+\alpha
_{2}\hat{\Omega}_{2}$ (with $\alpha _{1}$, $\alpha _{2}$ real). In general, $%
E_{1}$, $E_{2}$, $E$ will require three macroscopically-distinct
experimental arrangements, to respectively measure the non-commuting
observables $\hat{\Omega}_{1}$, $\hat{\Omega}_{2}$, $\alpha _{1}\hat{\Omega}%
_{1}+\alpha _{2}\hat{\Omega}_{2}$.

Now over an ensemble represented by some density operator $\hat{\rho}$,
quantum theory predicts that expectation values will be additive,%
\begin{equation*}
\langle \alpha _{1}\hat{\Omega}_{1}+\alpha _{2}\hat{\Omega}_{2}\rangle
=\alpha _{1}\langle \hat{\Omega}_{1}\rangle +\alpha _{2}\langle \hat{\Omega}%
_{2}\rangle
\end{equation*}%
the expectation for any $\hat{\Omega}$ being given by the Born rule%
\begin{equation}
\langle \hat{\Omega}\rangle =Tr(\hat{\rho}\hat{\Omega})  \label{Born}
\end{equation}

Indeed, for any number of non-commuting observables $\hat{\Omega}_{1}$, $%
\hat{\Omega}_{2}$, $\hat{\Omega}_{3}$, ....., and real numbers $\alpha _{1}$%
, $\alpha _{2}$, $\alpha _{3}$, ....., quantum theory predicts that%
\begin{equation}
\langle \alpha _{1}\hat{\Omega}_{1}+\alpha _{2}\hat{\Omega}_{2}+\alpha _{3}%
\hat{\Omega}_{3}+.....\rangle =\alpha _{1}\langle \hat{\Omega}_{1}\rangle
+\alpha _{2}\langle \hat{\Omega}_{2}\rangle +\alpha _{3}\langle \hat{\Omega}%
_{3}\rangle +.....  \label{add}
\end{equation}

In quantum theory, (\ref{add}) seems an inevitable and trivial consequence
of (\ref{Born}). But experimentally, it is a remarkable and highly
nontrivial feature of quantum theory that the statistics of quite \textit{%
distinct} experiments -- $E_{1}$, $E_{2}$, $E_{3}$, ....., $E$ -- should be
related in this simple and general way.

We shall see that for two-state systems expectation additivity is in fact
equivalent to the Born rule, and that both are violated in nonequilibrium.

\section{Two-State Systems}

Consider the quantum description of an arbitrary two-state system (or
two-dimensional subspace of any system). Every quantum observable $\hat{%
\Omega}$ may be written as a linear combination of the Pauli operators $\hat{%
\sigma}_{\mu }$ ($\mu =0$, $x$, $y$, $z$),%
\begin{equation*}
\hat{\Omega}=m_{\mu }\hat{\sigma}_{\mu }=m_{0}\hat{I}+\mathbf{m}\cdot 
\mathbf{\hat{\sigma}}
\end{equation*}%
(summing over repeated indices), where $\hat{\sigma}_{0}=\hat{I}$ is the
identity and the coefficients $m_{\mu }$ are real.

Observables of the form $\hat{\sigma}\equiv \mathbf{m}\cdot \mathbf{\hat{%
\sigma}}$, with $\mathbf{m}$ normalised to unity, might correspond to spin
(in units $\hbar /2$) along an axis $\mathbf{m}$ in real space. But in
general $\hat{\sigma}$ acts on any two-dimensional subspace for any system,
and $\mathbf{m}$ is a unit vector in an abstract 3-space (specifying a point 
$(\theta ,\phi )$ on the Bloch sphere). Whatever $\hat{\sigma}$ may
represent, quantum measurements of $\hat{\sigma}$ can yield outcomes $\sigma
=\pm 1$. No other outcomes are allowed. The distribution $p_{eq}^{\pm }(%
\mathbf{m})$ of outcomes in quantum equilibrium depends, of course, on the
quantum state.

For an ensemble with density operator $\hat{\rho}$, the (equilibrium) mean is%
\begin{equation*}
\left\langle \hat{\sigma}\right\rangle =Tr\left( \hat{\rho}\mathbf{m}\cdot 
\mathbf{\hat{\sigma}}\right) =\mathbf{m}\cdot Tr\left( \hat{\rho}\mathbf{%
\hat{\sigma}}\right) =\mathbf{m}\cdot \left\langle \mathbf{\hat{\sigma}}%
\right\rangle
\end{equation*}%
or%
\begin{equation*}
\left\langle \hat{\sigma}\right\rangle =\mathbf{m}\cdot \mathbf{P}=P\cos
\theta
\end{equation*}%
where the mean polarisation $\mathbf{P}=\langle \mathbf{\hat{\sigma}}\rangle 
$ (with norm $0\leq P\leq 1$) characterises the ensemble and $\theta $ is
the angle between $\mathbf{m}$ and $\mathbf{P}$.

With only two possible outcomes, the mean fixes the distribution:%
\begin{equation*}
p_{eq}^{\pm }(\mathbf{m})=\frac{1}{2}\left( 1\pm \langle \mathbf{\hat{\sigma}%
}\rangle \right)
\end{equation*}%
or%
\begin{equation}
p_{eq}^{\pm }(\mathbf{m})=\frac{1}{2}\left( 1\pm P\cos \theta \right)
\label{QTcos}
\end{equation}%
For a completely polarised beam, $P=1$ and $p_{eq}^{+}(\mathbf{m})=\cos
^{2}(\theta /2)$.

The equilibrium probabilities depend sinusoidally on $\theta $, for \textit{%
any} pure or mixed state (as long as it is not completely unpolarised). This
is a universal feature of any two-state quantum system (or qubit). As we
shall see, deviations from this behaviour are a hallmark of quantum
nonequilibrium.

\section{Quantum Probabilities from Additive Expectations}

It is remarkable that the quantum probabilities (\ref{QTcos}) are completely
determined by expectation additivity (\ref{add}).

For as an instance of (\ref{add}), consider axes $\mathbf{m}_{1}$, $\mathbf{m%
}_{2}$, $\mathbf{m}_{3}$ forming an orthonormal basis in the abstract
3-space. The observables $\mathbf{m}_{1}\cdot \mathbf{\hat{\sigma}}$, $%
\mathbf{m}_{2}\cdot \mathbf{\hat{\sigma}}$, $\mathbf{m}_{3}\cdot \mathbf{%
\hat{\sigma}}$ do not commute, and to measure them requires three distinct
experimental arrangements -- for example, three distinct orientations of a
Stern-Gerlach apparatus.

For an arbitrary unit vector%
\begin{equation*}
\mathbf{m}=c_{i}\mathbf{m}_{i}
\end{equation*}%
(summing over repeated indices, $i=1$, $2$, $3$) we have%
\begin{equation*}
\mathbf{m}\cdot \mathbf{\hat{\sigma}}=c_{i}\left( \mathbf{m}_{i}\cdot 
\mathbf{\hat{\sigma}}\right)
\end{equation*}%
Writing $E(\mathbf{m})\equiv \left\langle \mathbf{m}\cdot \mathbf{\hat{\sigma%
}}\right\rangle $, expectation additivity (\ref{add}) states that%
\begin{equation}
E(\mathbf{m})=c_{i}E(\mathbf{m}_{i})  \label{add2}
\end{equation}%
(regardless of the density operator $\hat{\rho}$ for the ensemble).

Now, under a change of (orthonormal) basis $\mathbf{m}_{i}\rightarrow 
\mathbf{m}_{i}^{\prime }$, the vector components $c_{i}$ transform as $%
c_{i}^{\prime }=R_{ij}c_{j}$, where $R_{ij}$ is an orthogonal rotation
matrix. For $E(\mathbf{m})$ to remain invariant for arbitrary $\mathbf{m}$
-- as it must, being a function of $\mathbf{m}$ and not of the basis used to
expand $\mathbf{m}$ -- the quantities $E(\mathbf{m}_{i})$ must also
transform like the components of a vector, $E(\mathbf{m}_{i}^{\prime
})=R_{ij}E(\mathbf{m}_{j})$. Thus%
\begin{equation*}
\mathbf{P}\equiv E(\mathbf{m}_{i})\mathbf{m}_{i}
\end{equation*}%
is a vector and we have%
\begin{equation}
E(\mathbf{m})=\mathbf{m}\cdot \mathbf{P\label{E}}
\end{equation}

Now $\left\vert E(\mathbf{m})\right\vert \leq 1$, so $0\leq P\leq 1$. Using
expectation additivity once more, we have $\mathbf{P}=\left\langle \mathbf{%
\hat{\sigma}}\right\rangle $ -- the mean polarisation of the ensemble. The
mean $E(\mathbf{m})$ again fixes the distribution, and so we obtain the
quantum probabilities (\ref{QTcos}).

We shall see that expectation additivity is very unnatural in
hidden-variables theories and peculiar to equilibrium.

\section{Nonequilibrium Systems}

We now consider hypothetical `nonequilibrium systems'. These are defined to
be systems where single outcomes (of quantum measurements) take only values
allowed by quantum theory, but where the statistical \textit{distribution}
of those values can disagree with quantum theory. For example, a spin-1/2
particle might still yield spin-measurement outcomes $\pm \hslash /2$, but
the statistical distribution of those outcomes over an ensemble might be
anomalous.

As we shall now show, such a nonequilibrium regime -- in which the space of
outcomes is the same and only the statistical distribution differs -- is a
natural extension of quantum physics in any deterministic hidden-variables
theory.\footnote{%
Of course, one might consider more complicated scenarios where individual
outcomes disagree with quantum theory. But then nothing general could be
said.}

For an ensemble of two-state systems with density operator $\hat{\rho}$,
quantum theory predicts a mean $\left\langle \hat{\sigma}\right\rangle =E(%
\mathbf{m})=\mathbf{m}\cdot \mathbf{P}$. In a deterministic hidden-variables
theory, the outcome $\sigma =\pm 1$ of a single quantum measurement of $\hat{%
\sigma}$ is determined by hidden parameters at $t=0$, collectively denoted $%
\lambda $. For a given measurement axis $\mathbf{m}$, there is a mapping%
\begin{equation}
\sigma =\sigma \left( \mathbf{m},\lambda \right)  \label{map1}
\end{equation}%
from the initial conditions $\lambda $ to the (unique) final outcome $\sigma 
$.

In principle, once (\ref{map1}) is given, the outcome of each quantum
experiment is determined by $\lambda $. Over an ensemble of similar
experiments, with fixed $\mathbf{m}$ and variable $\lambda $, there will be
a distribution of outcomes $\sigma \left( \mathbf{m},\lambda \right) $. For
an ensemble with a particular distribution $\rho _{eq}(\lambda )$ of hidden
variables, the distribution $p_{eq}^{\pm }(\mathbf{m})$ of outcomes will
agree with quantum theory,\footnote{%
A simple example of such a hidden-variables theory is given by Bell \cite%
{4,20,21}.} and the mean will be\footnote{%
As is customary, we write as if $\lambda $ were a continuous variable. The
integral sign is to be understood as a generalised sum, and no particular
assumption is being made about $\lambda $.}%
\begin{equation}
\left\langle \sigma \left( \mathbf{m},\lambda \right) \right\rangle
_{eq}\equiv \int d\lambda \ \rho _{eq}(\lambda )\sigma \left( \mathbf{m}%
,\lambda \right) =\mathbf{m}\cdot \mathbf{P}  \label{qumean}
\end{equation}

Let us now consider a `nonequilibrium ensemble' with distribution $\rho
(\lambda )\neq \rho _{eq}(\lambda )$. We retain the same deterministic
mapping (\ref{map1}) from $\lambda $ to outcomes, for each individual
system. In general, the distribution of outcomes over the ensemble will now 
\textit{disagree} with quantum theory, with a mean%
\begin{equation}
\left\langle \sigma \left( \mathbf{m},\lambda \right) \right\rangle \equiv
\int d\lambda \ \rho (\lambda )\sigma \left( \mathbf{m},\lambda \right) \neq 
\mathbf{m}\cdot \mathbf{P}  \label{nonqumean}
\end{equation}%
(or $E(\mathbf{m})\neq \mathbf{m}\cdot \mathbf{P}$) because the same
function $\sigma \left( \mathbf{m},\lambda \right) $ appears in (\ref{qumean}%
) and (\ref{nonqumean}).

It could happen that $\left\langle \sigma \left( \mathbf{m},\lambda \right)
\right\rangle =\left\langle \sigma \left( \mathbf{m},\lambda \right)
\right\rangle _{eq}$ for some special $\rho (\lambda )\neq \rho
_{eq}(\lambda )$. But in general the nonequilibrium mean will disagree with
the quantum mean, $\left\langle \sigma \left( \mathbf{m},\lambda \right)
\right\rangle \neq \left\langle \sigma \left( \mathbf{m},\lambda \right)
\right\rangle _{eq}$, and the nonequilibrium outcome probabilities%
\begin{equation*}
p^{\pm }(\mathbf{m})=\frac{1}{2}\left( 1\pm \left\langle \sigma \left( 
\mathbf{m},\lambda \right) \right\rangle \right)
\end{equation*}%
will disagree with the quantum values, $p^{\pm }(\mathbf{m})\neq p_{eq}^{\pm
}(\mathbf{m})$.

More formally, the set $S=\left\{ \lambda \right\} $ of values of $\lambda $
may be partitioned into%
\begin{equation*}
S^{+}(\mathbf{m})=\left\{ \lambda |\sigma \left( \mathbf{m},\lambda \right)
=+1\right\} ,\;\;\;\;\;S^{-}(\mathbf{m})=\left\{ \lambda |\sigma \left( 
\mathbf{m},\lambda \right) =-1\right\}
\end{equation*}%
where the sets $S^{\pm }(\mathbf{m})$ are fixed by the mapping (\ref{map1})
and are therefore \textit{independent} of the ensemble distribution $\rho
(\lambda )$. For the equilibrium measure $d\mu _{eq}\equiv \rho
_{eq}(\lambda )d\lambda $, we have $\mu _{eq}[S^{\pm }(\mathbf{m}%
)]=p_{eq}^{\pm }(\mathbf{m})$. But for an arbitrary nonequilibrium measure $%
d\mu \equiv \rho (\lambda )d\lambda $, in general $\mu \lbrack S^{\pm }(%
\mathbf{m})]\neq \mu _{eq}[S^{\pm }(\mathbf{m})]$ and $p^{\pm }(\mathbf{m}%
)\neq p_{eq}^{\pm }(\mathbf{m})$.\footnote{%
Of course, for a given $\mathbf{m}$ there are special $\rho (\lambda )\neq
\rho _{eq}(\lambda )$ such that $\mu \lbrack S^{\pm }(\mathbf{m})]$ happens
to equal $\mu _{eq}[S^{\pm }(\mathbf{m})]$, but generically $\mu \lbrack
S^{\pm }(\mathbf{m})]\neq \mu _{eq}[S^{\pm }(\mathbf{m})]$.}

Thus, nonequilibrium will generically break the sinusoidal probability
modulation (\ref{QTcos}) predicted by quantum theory for any two-state
system. Equivalently, in general $E(\mathbf{m})\neq \mathbf{m}\cdot \mathbf{P%
}$ and expectations are non-additive.

Note the key conceptual point: we have the same deterministic mapping (\ref%
{map1}) for each system, regardless of the ensemble distribution.

For pairs of entangled two-state systems, the correct quantum correlations
will be obtained, for some ensemble distribution $\rho _{eq}(\lambda )$,
only if the outcomes for at least one subspace depend on the measurement
setting for the other (as shown by Bell \cite{4,20}). This nonlocality, or
more generally `contextuality', appears in the deterministic mapping from $%
\lambda $ to outcomes. For a nonequilibrium ensemble $\rho (\lambda )\neq
\rho _{eq}(\lambda )$, the marginal \textit{statistics} for at least one
subspace generally depend on what measurements are performed for the other 
\cite{15,16,22}. Such nonlocal or contextual statistics are discussed
elsewhere as a signature of non-quantum systems \cite{22}. Here, we focus on
the simpler signature of non-additive expectations for a single two-state
system.

Finally, note that the pilot-wave formulation of quantum theory [1--10]
provides a specific and generally-applicable hidden-variables theory that
may readily be extended to nonequilibrium ensembles [5, 10--14, 18, 19].

In pilot-wave theory, a system (of particles or fields) with wave function $%
\psi (x,t)$ has a configuration $x(t)$ whose motion is given by $\dot{x}%
(t)=j(x,t)/\left\vert \psi (x,t)\right\vert ^{2}$ -- where in quantum theory 
$j=j\left[ \psi \right] =j(x,t)$ is called the `probability current'%
\footnote{%
For example, for a single spinless particle of mass $m$, $j=(\hbar /m)\func{%
Im}\left( \psi ^{\ast }\nabla \psi \right) $.} but where in pilot-wave
theory $\psi $ is regarded as an objective physical field (in configuration
space) guiding the system.

This dynamics for single systems yields the correct distribution of quantum
outcomes over ensembles, \textit{if} the configurations -- or hidden
variables\footnote{%
Actually, a complete specification of $\lambda $ here is $x_{0}$ and $\psi
_{0}(x)$.} -- at $t=0$ have the equilibrium distribution $\rho
_{0}(x)=\left\vert \psi _{0}(x)\right\vert ^{2}$ (which implies $\rho
(x,t)=\left\vert \psi (x,t)\right\vert ^{2}$ for all $t$). But one can just
as well consider an initial ensemble $\rho _{0}(x)\neq \left\vert \psi
_{0}(x)\right\vert ^{2}$, whose evolution is given by%
\begin{equation}
\frac{\partial \rho }{\partial t}+\nabla \cdot \left( \rho \dot{x}\right) =0
\label{cont}
\end{equation}%
In appropriate circumstances, (\ref{cont}) leads to relaxation $\rho
\rightarrow \left\vert \psi \right\vert ^{2}$ on a coarse-grained level \cite%
{5,10,11,14}, much as the corresponding classical evolution on phase space
leads to thermal relaxation. But for as long as the ensemble remains in
nonequilibrium, the statistics of outcomes of quantum measurements will
disagree with quantum theory.

\section{Non-Quantum Probabilities and Non-Additive Expectations}

We have seen that quantum nonequilibrium generically breaks the sinusoidal
probability modulation (\ref{QTcos}) predicted by quantum theory for
two-state systems. In general, $E(\mathbf{m})\neq \mathbf{m}\cdot \mathbf{P}$
and expectation additivity (\ref{add2}) is violated.

Some insight into why (\ref{add2}) is violated can be obtained by
considering an `extreme' nonequilibrium ensemble with $\rho (\lambda )$
concentrated on just one value $\lambda =\lambda _{0}$. Such an ensemble has
zero dispersion for all quantum observables, the outcome of any measurement
being determined by $\lambda _{0}$ (together with the experimental
settings). Thus the mean of $\sigma \left( \mathbf{m},\lambda \right) $ is
simply%
\begin{equation*}
\left\langle \sigma \left( \mathbf{m},\lambda \right) \right\rangle =\int
d\lambda \ \rho (\lambda )\sigma \left( \mathbf{m},\lambda \right) =\sigma
\left( \mathbf{m},\lambda _{0}\right)
\end{equation*}%
for any measurement setting $\mathbf{m}$. Taking (for simplicity) $c_{3}=0$,
(\ref{add2}) then reads%
\begin{equation}
\sigma \left( \mathbf{m},\lambda _{0}\right) =c_{1}\sigma \left( \mathbf{m}%
_{1},\lambda _{0}\right) +c_{2}\sigma \left( \mathbf{m}_{2},\lambda
_{0}\right)  \label{contra}
\end{equation}%
This can hold only if $\left\vert c_{1}+c_{2}\right\vert $ or $\left\vert
c_{1}-c_{2}\right\vert $ equals $1$, which requires $c_{1}$ or $c_{2}$ to
vanish. If $c_{1}c_{2}\neq 0$, (\ref{contra}) cannot be satisfied for any $%
\lambda _{0}$.\footnote{%
This point was made by Bell \cite{4,21}, in his refutation of von Neumann's
supposed proof \cite{23} that hidden variables are incompatible with quantum
theory.}

This example makes it clear that expectation additivity is an exceptional
feature, peculiar to equilibrium. For \textit{all} $\lambda $,%
\begin{equation*}
\sigma \left( \mathbf{m},\lambda \right) \neq c_{1}\sigma \left( \mathbf{m}%
_{1},\lambda \right) +c_{2}\sigma \left( \mathbf{m}_{2},\lambda \right)
\end{equation*}%
(if $c_{1}c_{2}\neq 0$), and yet in equilibrium%
\begin{equation*}
\left\langle \sigma \left( \mathbf{m},\lambda \right) \right\rangle
_{eq}=c_{1}\left\langle \sigma \left( \mathbf{m}_{1},\lambda \right)
\right\rangle _{eq}+c_{2}\left\langle \sigma \left( \mathbf{m}_{2},\lambda
\right) \right\rangle _{eq}
\end{equation*}%
Quantities that are always unequal for individual systems turn out to be
equal when averaged over an equilibrium ensemble $\rho _{eq}(\lambda )$.

Expectation additivity for incompatible experiments is completely unnatural
-- and generically false -- in any hidden-variables theory. It is, as it
were, a `conspiracy' of the equilibrium state. The associated two-state
probabilities (\ref{QTcos}) are equally unnatural and exceptional.

\section{Experimental Tests}

The above signature could be searched for experimentally in any system. Since%
\begin{equation*}
p^{+}(\mathbf{m})=\frac{1}{2}\left( 1+\left\langle \sigma \left( \mathbf{m}%
,\lambda \right) \right\rangle \right)
\end{equation*}%
a test of how $p^{+}(\mathbf{m})$ varies with $\mathbf{m}$ is also a test of
how the expectation $E(\mathbf{m})=\left\langle \sigma \left( \mathbf{m}%
,\lambda \right) \right\rangle $ varies with $\mathbf{m}$. And because $E(%
\mathbf{m})=\mathbf{m}\cdot \mathbf{P}$ is equivalent to expectation
additivity, any test of the former is a test of the latter.

Consider photons, whose polarisation forms a two-state system. The values $%
\sigma =\pm 1$ of the observable $\hat{\sigma}=\mathbf{m}\cdot \mathbf{\hat{%
\sigma}}$ now correspond respectively to polarisation parallel or
perpendicular to an axis $\mathbf{M}$ in 3-space -- where an angle $\theta $
on the Bloch sphere corresponds to an angle $\Theta =\theta /2$ in real
space.

If a beam of single photons with density operator $\hat{\rho}$ is incident
on a linear polariser set at angle $\Theta $ (relative to the axis of
maximum transmission), quantum theory predicts that the fraction of photons
transmitted will be (from (\ref{QTcos}))%
\begin{equation}
p^{+}(\Theta )=p_{eq}^{+}(\Theta )=\frac{1}{2}\left( 1+P\cos 2\Theta \right)
\label{Photcos}
\end{equation}%
where $0\leq P\leq 1$ is characteristic of the beam. For a fully-polarised
beam $P=1$, and $p_{eq}^{+}(\Theta )=\cos ^{2}\Theta $.

The equilibrium transmission probability (\ref{Photcos}) depends
sinusoidally on $\Theta $ for any incident beam (with a well-defined $\hat{%
\rho}$) that is not completely unpolarised. From a hidden-variables
perspective this is an exceptional feature of equilibrium, reflecting
additivity of expectations for incompatible experiments -- an additivity
that is satisfied for equilibrium ensembles but not in general. Thus,
deviations from (\ref{Photcos}) would provide a distinctive signature of
nonequilibrium.

We therefore propose that experiments be performed to search for quantum
nonequilibrium by testing polarisation probabilities for single photons.
Deviations from (\ref{Photcos}) would provide an unambiguous violation of
quantum theory.

Photons from very distant sources seem of particular interest. For it has
been argued that, as the quantum probability distribution for a single
photon spreads out over cosmological distances, any small-scale deviations
from equilibrium could be stretched up to larger lengthscales \cite{13,14}.

Incoming photons may be filtered or prepared (for example in a pure
polarisation state) before being subjected to further measurements. As long
as the prepared ensemble has a well-defined density operator $\hat{\rho}$,
quantum theory predicts a sinusoidal modulation of $p^{+}(\Theta )$ (for any 
$\hat{\rho}$ with $P\neq 0$).

One might wish to dispense with preparing the incoming photons, to minimise
any possible disturbance of the sought-for nonequilibrium. There are of
course many astrophysical sources of polarised photons, involving scattering
or synchrotron emission, at cosmological distances. For example, strong ($%
\sim 80\%$) linear polarisation was recently seen in single photons from a
gamma-ray burst \cite{24}, and the cosmic microwave background is slightly
polarised \cite{25,26}. But without a controlled preparation, the received
photons need not have a well-defined density operator $\hat{\rho}$ or mean
polarisation $\mathbf{P}$. Though one could try to learn about and
characterise the source by appropriate measurements.

An open question is whether one could dispense with a preparation without
knowing anything at all about the source. Given a large number $N$ of single
photons, polarisations along an angle $\Theta $ may be measured for a large
sample drawn at random. One might expect that in the limit of large $N$ and
large samples, any quantum source would yield a mean of the form $\mathbf{m}%
\cdot \mathbf{P}$ (where $P$ could be zero). This remains to be studied. If
true, there would be no need to prepare the photons or to know anything
about the source. In practice, the incoming photons form a time ensemble. A
random sample could be drawn from the whole ensemble, for each $\Theta $, by
resetting $\Theta $ randomly for each photon. Data sets with the same $%
\Theta $ would provide values of $p^{+}(\Theta )$, to be compared with (\ref%
{Photcos}).

We emphasise that, with a controlled filtering or preparation of the
incoming photon polarisations, corresponding to a well-defined density
operator $\hat{\rho}$, the quantum prediction (\ref{Photcos}) is unambiguous
and the observation of a non-sinusoidal modulation would signal a violation
of quantum theory.

An especially attractive possibility would arise if relic gravitons were
detected from the very early universe \cite{27} and if their polarisation
could be studied. And of course, polarisation probabilities could be tested
in the laboratory for any kind of particle at high energies.

\textbf{Acknowledgements.} I am grateful to David Buscher, Wayne Myrvold,
Itamar Pitowsky, David Poulin and Robert Spekkens for helpful comments, and
to Howard Burton and Lee Smolin for their support and encouragement.

\end{document}